# A strategy for fabricating micro-scale freestanding single-crystalline complex oxide device arrays

Huandong Chen[a], Yang Liu[a], Harish Kumarasubramanian[a], Mythili Surendran[a,c], Jayakanth Ravichandran[a,b,c]*

[a]Mork Family Department of Chemical Engineering and Materials Science, University of Southern California, Los Angeles, CA 90089, USA

[b]Ming Hsieh Department of Electrical and Computer Engineering, University of Southern California, Los Angeles, CA 90089, USA

[c]Core Center for Excellence in Nano Imaging, University of Southern California, Los Angeles, CA 90089, USA

*Email: j.ravichandran@usc.edu

Keywords: complex oxides; epitaxial lift-off; BCZT; device fabrication; ferroelectric capacitors

**Abstract**

We present a general fabrication strategy for freestanding single-crystalline complex oxide device arrays *via* wet chemical etching-based microfabrication processes and epitaxial lift-off. Here, we used $0.5Ba(Zr_{0.2}Ti_{0.8})O_3$-$0.5Ba(Zr_{0.7}Ti_{0.3})O_3$ (BCZT) as a model relaxor ferroelectric oxide system and $La_{0.7}Sr_{0.3}MnO_3$ as the sacrificial layer for demonstration. Arrays of $SrRuO_3$ (SRO) / BCZT / SRO ferroelectric capacitor mesas were first defined and isolated on the growth wafer, and then they were released using epitaxial lift-off with lithography-defined surrounding etching holes, after which the freestanding device arrays were integrated onto a glass substrate. Our proposed strategy sheds light on preparing various freestanding single-crystalline oxide devices and paves the way for their heterogeneous integration onto arbitrary substates.

# Introduction

Epitaxial lift-off (ELO) is a technological process designed to obtain freestanding single crystalline films by releasing thin epitaxial layers from their original growth substrates[1]. This technique has been successfully employed on III-V materials over the past 30 years[2-4] to produce their heterogeneously integrated electronics and optoelectronics[5-7] that feature device characteristics not possible with film-on-substrate geometry, novel mechanical and electronic functionalities, and reduced costs. Epitaxially grown complex oxides belong to an interesting class of materials that have shown rich physical and functional properties ranging from ferroelectricity[8-10], high mobility electron-gas[11], superconductivity[12-14], to engineered interface phenomena[11, 15, 16], however, the extended applications of these exciting phenomena have been limited by the form of rigid films that directly grow on single crystalline oxide substrates for a long time. Inspired by the success of freestanding III-V device applications[2, 5] and the proliferation of research on mechanically stacked 2D heterostructures[17, 18], recently, there is a rapid increase of interests in developing ELO-based processes for obtaining freestanding single-crystalline complex oxides and further achieving their heterogeneous integration[19-22]. In this experimental scheme, an epitaxially-grown, sacrificial layer ($La_{0.7}Sr_{0.3}MnO_3$ (LSMO)[19, 20] or $Sr_3Al_2O_6$ (SAO)[21, 22]) is inserted between the oxide substrate and active oxide layers[21-23] during the growth and is then laterally etched to release the entire top stack, which is then transferred and integrated on various desired materials or substrates[19, 20].

Following several pioneering work in the mid 2010s[19, 21], a wide variety of freestanding single crystalline oxide films have been obtained and heterogeneously integrated with flexible substrates or semiconductors, enabling several unique applications such as flexible oxide electronics[20, 24], high-$\kappa$ gating dielectric integrations[25, 26] and extreme strain modulations[27, 28] that cannot be achieved with as-grown complex oxides. Nonetheless, most of current research centers on either obtaining the entire oxide membrane[29-31] or unitizing a single broken piece of micro-flake for material characterization and prototype device demonstration[26, 32]. The large-scale-compatible fabrication procedures of single-crystalline freestanding complex oxide devices or device arrays remain largely underexplored, compared to well-established III-V materials.

Here, using relaxor ferroelectric $0.5Ba(Zr_{0.2}Ti_{0.8})O_3$-$0.5Ba(Zr_{0.7}Ti_{0.3})O_3$ (BCZT) as a model complex oxide system, $SrRuO_3$ (SRO) as both top and bottom electrodes, and LSMO as the oxide sacrificial layer, we present detailed microfabrication procedures of obtaining freestanding, isolated SRO/BCZT/SRO capacitor arrays and their heterogeneous integration. $10 \times 10$ mesas of BCZT metal-insulator-metal (MIM) capacitor arrays were first individually defined and fabricated on the as-grown oxide wafer using standard photolithography and optimized wet chemical etching processes. Etch holes were implemented surrounding each device for lift-off, during which the LSMO etchant was delivered to the sacrificial layer through these etch holes in a controlled manner while the sidewalls of the active device stack remained well-protected. Moreover, the freestanding device arrays were retrieved either individually or as a whole from the growth substrate and integrated onto an alien substrate using a UV-curable adhesive. Our strategy, when combined with appropriate etching protocols, is generally applicable for fabricating many other micro-scale freestanding complex oxide devices or device arrays, which adds more exciting possibilities to the current oxide membrane research.

**Epitaxial growth of releasable SRO/BCZT/SRO stack**

BCZT, a solid solution of $BaTiO_3$, is a lead-free relaxor ferroelectric that features high room temperature dielectric constant[33], large piezoelectric coefficient[33, 34], and strong electro-optic (EO) coupling[35], making it promising for various applications such as energy storage[36], actuators[33], and EO modulators[35]. Recently, a sub-group of the authors reported the epitaxial growth of high quality BCZT thin films using pulsed laser deposition (PLD)[37].

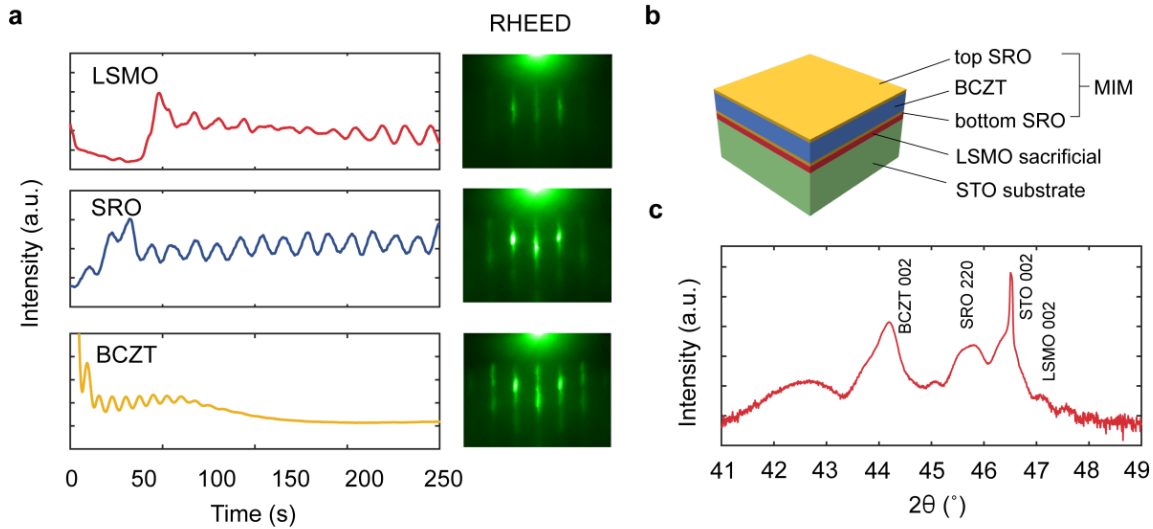

**Figure 1** (**a**) Representative oscillations in the RHEED specular spot intensity and diffraction patterns of LSMO (top), SRO (middle), and BCZT (bottom) layers during the growth. (**b**) Schematic illustration of the releasable epitaxial SRO/BCZT/SRO device stack. (**c**) $\theta$-$2\theta$ scan of an as-grown releasable BCZT epitaxial device stack.

As illustrated in Figure 1a and 1b, the growth of the releasable BCZT device stack used in this work started with the formation of a 20 nm LSMO sacrificial layer on a TiO$_2$-terminated SrTiO$_3$ (STO) substrate, followed by a sequential deposition of 20 nm SRO bottom electrode layer, a 100 nm BCZT relaxor ferroelectric layer, and a 20 nm SRO top electrode layer in the same chamber. Optimal growth conditions were applied for each layer and reflection high energy electron diffraction (RHEED) was used to monitor the film quality and growth rate *in situ*. High quality SRO ($a_{SRO}$ = 3.93 Å) and BCZT ($a_{BCZT}$ = 4.00 Å) films were obtained on LSMO with layer-by-layer growth, as revealed by the RHEED patterns and oscillations in the intensity of the specular spot as shown in Figure 1a. It is noted that LSMO has an in-plane lattice parameter of 3.87 Å, which is close to that of both STO substrate ($a_{STO}$ = 3.91 Å) and SRO layer, and hence the lattice mismatch is of ~ 1.5%. Figure 1c illustrates a representative $\theta$-$2\theta$ scan of the as-grown MIM epitaxial device stack using high-resolution thin film X-ray diffraction (XRD). The reflections of LSMO 002, SRO 220 and BCZT 002 were clearly observed, indicating a good epitaxial growth of each layer on the STO substrate.

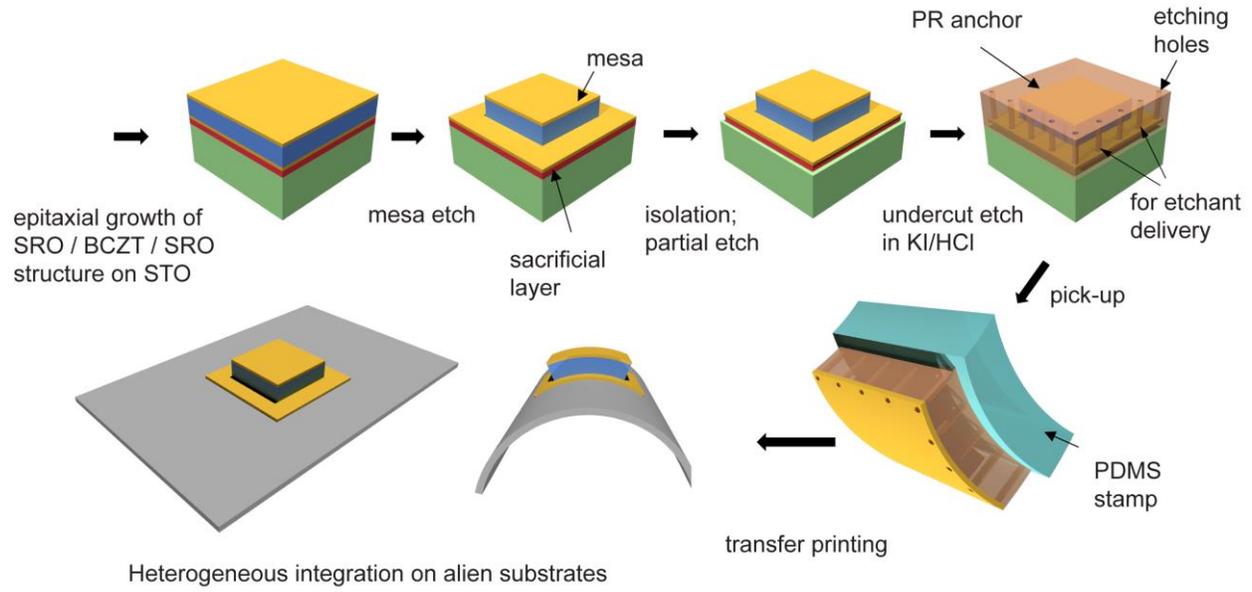

**Figure 2** Schematic illustration of fabrication processes for micro-scale freestanding SRO/BCZT/SRO device arrays.

In this experimental scheme, epitaxially grown conductive oxide SRO was employed as both top and bottom electrodes of the BCZT MIM capacitors to improve the metal-to-ferroelectric interface quality. Recently, such all-oxide MIM architecture has shown great promise in achieving excellent device performance such as low switching voltage and high endurance, as demonstrated in another archetypal perovskite ferroelectric $BaTiO_3$[38, 39].

## Fabrication of micro-scale, isolated all-oxide BCZT MIM capacitors

The entire micro-fabrication procedures for freestanding SRO/BCZT/SRO capacitors include the following three major steps: 1) fabrication of lithography-defined BCZT MIM capacitor arrays on wafer through wet chemical etching, 2) releasing of as-fabricated devices *via* epitaxial lift-off through surrounding etch holes, and 3) heterogeneous integration of freestanding device arrays on desired alien substrates, as illustrated in Figure 2.

The on-wafer fabrication processes of all-oxide BCZT MIM capacitor arrays started with the definition of mesa-shaped active device arrays (10 × 10, ~100 × 100 μm$^2$ each) using regular photolithography, followed by successive wet chemical etching of top SRO layer and BCZT layer (Figure 3a, bottom left and bottom right). A chemical etchant based on sodium periodate (0.4 mol/L NaIO$_4$ in DI) was used for removing SRO, which has been reported effective with an intermediate etch rate of ~ 2 nm/s[40]. It is important to note that the NaIO$_4$-based etchant removes SRO exclusively by oxidizing Ru species, which leads to a perfect etch stop at BCZT surface. A slight color change of the sample was noticed when comparing between the optical microscopic images taken before (Figure 3a, top right) and after SRO etching (Figure 3a, bottom left), which has been adopted to monitor the etching process. An etching step of ~ 20 s in NaIO$_4$ solution is typically sufficient to remove the SRO layer and no further discernible change in color can be observed even if the sample was left over a much longer time (e.g., ~ 5 min) in the etchant.

On the other hand, the etchant for BCZT film has not been well established in the literature. Nonetheless, inspired by the chemical etchants reported for processing BaTiO$_3$, Ba$_{0.5}$Sr$_{0.5}$TiO$_3$, and Pb(Zr$_x$Ti$_{1-x}$)O$_3$-based devices[41, 42], which is usually a mixture of buffered hydrofluoric acid (BHF) and acids such as HNO$_3$, H$_2$SO$_4$, HCl and H$_3$PO$_4$, we optimized the BCZT etchant and etching processes based on BHF and HCl. As shown in Figure S5 and S6, as-received BHF etchant (BOE 7:1) easily leaves significant etching residues despite its effectiveness in removing BCZT, while HCl etchant (HCl : DI = 1 :4, in volume) itself suffers from a slow etch rate, which is likely limited by the insufficient removal of Zr species. Therefore, the optimized BCZT etchant used for this study composes of both BHF and HCl (BHF : HCl : DI = 1 : 10 : 40, in volume), where small amount of HF is added to a diluted aqueous solution of HCl to effectively remove zirconium species, and hence, continuous etching of BCZT is achieved. Etching residues are minimized by gently swirling the container during the wet etching process and ~ 1 nm/s etch rate of BCZT film is typically obtained. An obvious and gradual color change was observed by bare eyes during the BCZT wet etching due to the thinning of the BCZT dielectric layer, which confirms the continuous removal of BCZT and has been used to monitor the etching progress. Figure 3a (bottom right) shows an optical microscopic image of the sample after the BCZT etching step, with the bottom SRO layer exposed. It is important to note that the bottom SRO layer also acts as a good etch stop for this optimized BCZT etchant, as evidenced by the electrical current-voltage (*I-V*) measurements (Figure 3b). A two-probe resistance of ~ 1 kΩ was observed through direct tungsten probes on exposed oxide surfaces after BCZT etching, while it remains insulating before the etching. Therefore, controlled wet etching processing of SRO/BCZT/SRO device stack with an excellent etching selectivity has been achieved.

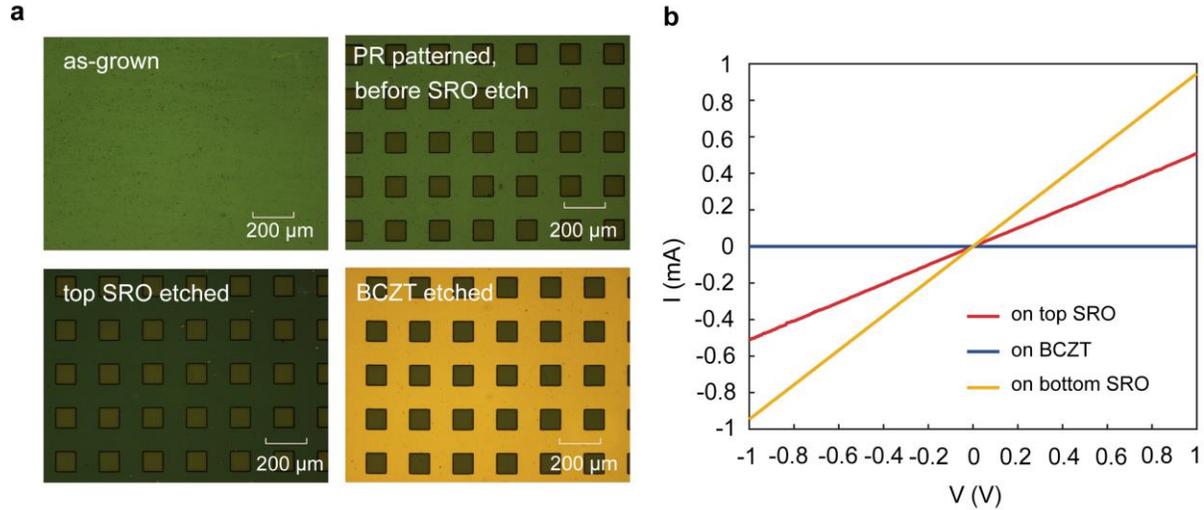

**Figure 3** (**a**) Optical micrographic images of the mesa etch processes on wafer. Top SRO and BCZT layers were chemically etched sequentially (bottom left and bottom right) to form active regions of the SRO/BCZT/SRO device arrays. (**b**) Representative current-voltage (*I-V*) characteristics of exposed top surfaces during etching. The stark contrast of electrical conductivity between metallic SRO and insulating BCZT has been used to monitor etching progresses.

Representative *I-V* and *P-V* characteristics of such an SRO/BCZT/SRO ferroelectric capacitor fabricated on wafer are illustrated in Figure S7. A low switching voltage of <0.4 V was clearly observed in the system, consistent with previous reports on SRO/BTO/SRO-based ferroelectric systems[38, 39]. Moreover, the switching curve is more symmetric compared to previously reported SRO/BCZT/Ti/Au[37], thanks to the symmetric SRO electrode design.

**Epitaxial lift-off and heterogeneous integration**

To release as-fabricated BCZT device arrays from the growth substrate, an epitaxial lift-off process with pre-defined surrounding etching holes was adopted. Similar strategy has been widely employed to fabricate freestanding III-V devices[6, 7, 43, 44]; however, to the best of our knowledge, no such attempt has been made on complex oxides. Isolated arrays of individual BCZT capacitors (~180 × 180 µm$^2$) were first defined by photolithography and wet etching of the bottom SRO layer, followed by a partial etching step of the LSMO sacrificial layer in a diluted solution based on potassium iodide (KI) and HCl (Figure 4a) and subsequent photoresist etch holes formation (Figure 4b, left). It is noted that the channels for delivering the etchant to the LSMO sacrificial layer (etching holes) were made outside the active device area to minimize the exposure of BCZT sidewalls to the etchant such that undesired chemical etching-related degradation of the active devices can be minimized, as illustrated in Figure S3. Here, an "*L*-shaped" etching holes arrangement design was employed for a proof-of-concept demonstration for complex oxides, which was originally optimized for III-V processing and found effective in mitigating the fracture at the cell centers due to stress accumulation, as reported in a study of freestanding triple-junction

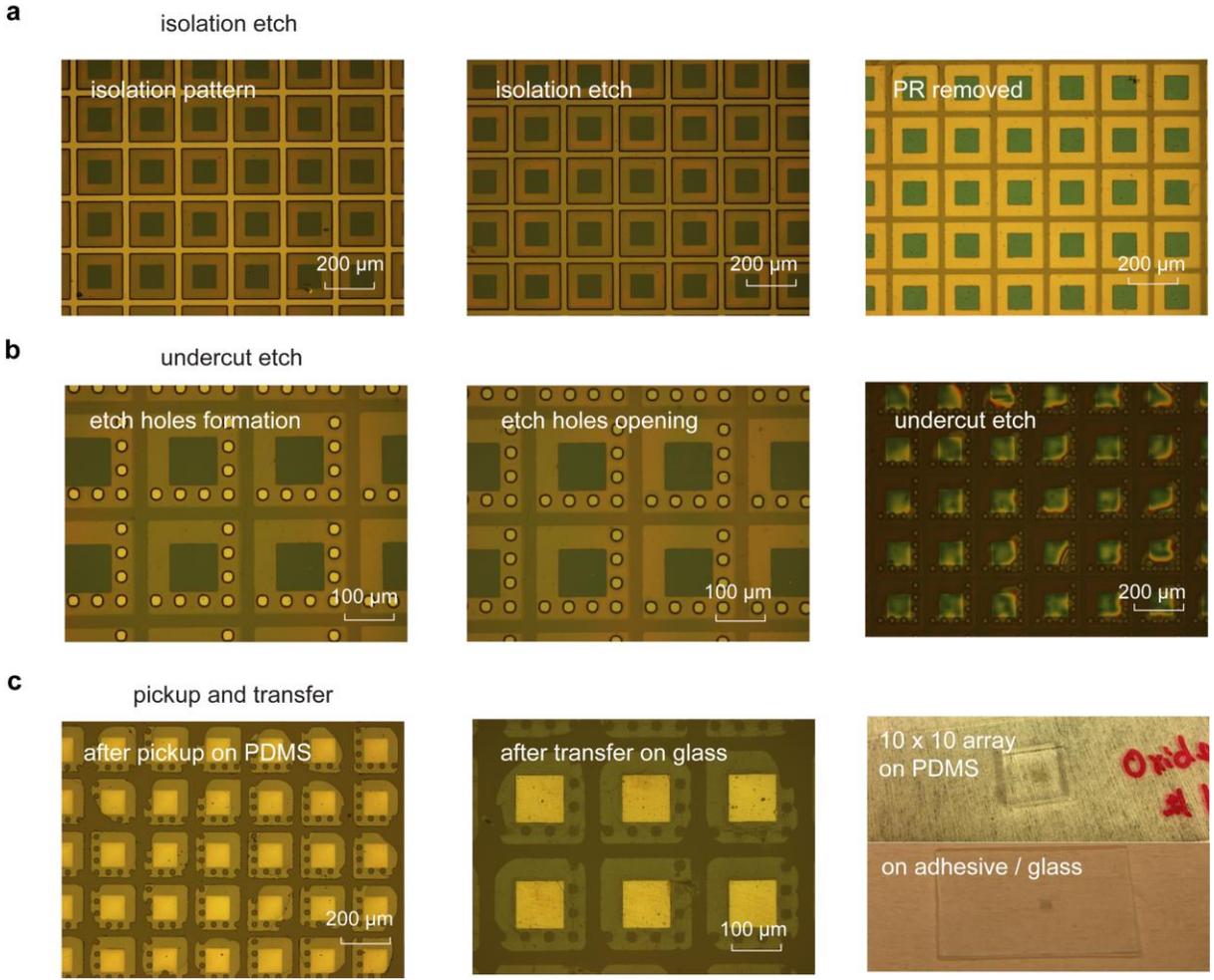

**Figure 4** (**a**) Optical micrographs of SRO/BCZT/SRO device arrays after isolation patterning (left), isolation etching (middle), and PR removal (right). (**b**) Optical micrographic images of the devices after "L-shaped" etching holes formation (left), etching holes opening (middle), and undercut etching (right). (**c**) Optical micrographic and photographic images of freestanding SRO/BCZT/SRO device arrays after picked up by PDMS (left and right top) and transferred on glass (middle and right bottom).

III-V solar cells by Gai *et al*[43]. In general, other surrounding etching hole designs with a different size/density/distribution would also serve the purpose of individual device releasing, although the overall etching time and device yield can be slightly affected.

After patterning the etching holes structure with photoresist, an additional SRO etching step was applied to "open" the etching holes by removing the bottom SRO layer in exposed regions (Figure 4b, middle). The undercut etching process was carried out by simply dipping the whole sample into the HCl/KI-based LSMO etchant[19, 20]. The etchant was delivered to each individual device through the surrounding etching holes simultaneously, which greatly reduces the overall undercut etching time, and it typically takes 3 to 4 hours to release devices with ~ 200 µm lateral dimensions (Figure S8). The chances of crack formation are relatively low in those devices during undercut etching and a high yield of 97/100 (without visible film cracks) was achieved, as illustrated in Figure 4b (right). This is primarily attributed to the improved mechanical support from photoresist "anchoring" and controlled etchant delivery through etching holes.

The freestanding BCZT device arrays were then picked up from the growth wafer using an elastomeric PDMS stamp and transferred onto a glass substrate, following the transfer printing procedures reported elsewhere[7, 43, 45], as shown in Figure 4c and S9. Here, a thin spin-on-glass-based UV-curable adhesive (~ 1 µm thick) was employed to facilitate the transfer process. Glass was chosen in this study for the proof-of-concept demonstration (Figure 4c) and other substrates such as silicon, flexible PET or polyimide sheets can be readily used depending on the needs of the actual application. An adhesive-free transfer process might need to be developed in the future for scenarios when the oxide-substrate interface is critical. For instance, a surface tension force-assisted van der Waals bonding method has been demonstrated for integrating freestanding GaAs films onto arbitrary substrates[46], which can be useful for developing similar processes for complex oxides. It is noted that ~20 more devices got cracked after pickup and transferring (Figure S10), which might be attributed to the over pressing of the PDMS stamp during the transfer process. Additional optimization of processing conditions would further improve the overall yield in the future.

**Discussion and Prospects**

On the other hand, so far, electrical characterization of transferred BCZT capacitor arrays remains challenging with the current design of the device configuration, i.e., ~ 100 nm-thick relaxor ferroelectric layer with a 20 nm-thick wing-shaped bottom SRO electrode (Figure 4c, middle) on a "soft" adhesive. Our attempts to directly probe SRO electrodes using tungsten tips in a probe station was not successful due to mechanical damage created on the transferred devices by the probes. To resolve this issue, one may need to explore a proper packaging route to electrically contact those device arrays in the future, including polymer encapsulation, VIA (vertical interconnect access) formation and metal interconnection, as has been widely applied in modern microchip manufacturing[47] and many studies on flexible and stretchable electronics[45, 48, 49]. A schematic of such device packing processes is illustrated in Figure S11.

Nonetheless, the proposed strategy and device fabrications procedures in this work are not only compatible to large-area device processing and packaging, but also generally applicable to many other commonly used complex oxide systems such as $BaTiO_3$, $PbZr_xTi_{1-x}O_3$, and $PbTiO_3$/STO superlattices with minimal modifications in the specifications of the processes, especially etching proctols. Further, instead of using wet chemical etching that may lead to tapered sidewall profile, the anisotropic dry etching methods such as inductively coupled plasma-reactive ion etching can be applied to accurately define the active device shape (mesa etch step) and maintain vertical sidewalls. Vertical sidewalls are critical in developing freestanding optical devices such as waveguides and electro-optical modulators, as their device performance can be largely impacted by the distortion of device geometry, although the etching selectively can be largely compromised for physical bombardment processes. In the meantime, the etching holes-based releasing strategy still applies in this case.

Moreover, this strategy can be extended to systems using a water-soluble sacrificial layer such as SAO, which have been more popular than LSMO for obtaining freestanding oxide membranes over the years due to its simplicity on film release. The epitaxial lift-off strategy for oxide device arrays presented in this work is in principle compatible with those systems as well, with the potential for unnecessary sacrificial layer lateral etching during the device wet chemical processes such as photoresist developing and wet etching. For water-soluble layers that feature fast releasing such as $Sr_4Al_2O_7$[50], additional care needs to be taken when designing the on-wafer device processes to minimize the lateral etching, while for most of reported $Sr_3Al_2O_6$ layers that require a few hours or days to release the oxide membranes[21, 51], such effect of lateral etching is expected to be minimal.

## Conclusion

In conclusion, we have successfully fabricated freestanding capacitor device arrays of SRO/BCZT/SRO using wet chemical etching-based device fabrication processes and controlled epitaxial lift-off through etching holes. Our work sheds light on the heterogeneous integration of freestanding complex oxides for device applications.

## Methods:

**Epitaxial thin film growth and structural analysis**

A TSST PLD system with a 248 nm KrF excimer laser was used to deposit the BCZT device stack on either $TiO_2$-terminated STO or $GdSrO_3$ (GSO) substrate. LSMO, SRO, and BCZT films were grown under the same growth temperature of 750˚C with fixed laser fluence of 1.5 J/cm$^2$, while the growth oxygen pressure was tuned accordingly to optimize the film quality (LSMO: 20 mbar,

SRO: 100 mbar, and BCZT: 10 mbar). The device stack was then cooled to room temperature at a rate of 5˚C/min after growth.

The high-resolution thin film out-of-plane X-ray diffraction scan was performed in a Bruker D8 Advance diffractometer using a Ge (004) two bounce monochromator with Cu radiation at room temperature.

A BCZT/LSMO/GSO sample was prepared separately and the entire BCZT film was released following a conventional undercut etching and polymer stamping method[21] to characterize the quality of transferred BCZT film. A comparison of XRD scans before and after transfer are illustrated in Figure S1 and Figure S2 shows piezoelectric force microscopy (PFM) characterization of a transferred BCZT film.

**Device fabrication**

Regular photolithography was employed to form all the required patterns with AZ1518 positive photoresist (Merck KGaA, with 110˚C, 3 min hard bake before etching) and MJB3 (Karl Suss) mask aligner for each wet chemical etching step. Samples with only 1-2 mm lateral dimensions (broken from a 5 mm × 5 mm as grown substrate) were used for wet etching condition optimization. Additional care has been taken to handle photolithography processes on those small samples to minimize the edge bead formation, as illustrated in Figure S4. With the use of supporting substrates for spin-coating, relatively flat photoresist films can be obtained.

0.4 mol/L $NaIO_4$ (>99.8%, Sigma Aldrich) in DI was used as the SRO etchant. The optimized BCZT etchant (BHF : HCl : DI = 1 : 10 : 40) was prepared by first preparing a mixture of HCl (36.5-38 %, VWR Chemicals BDH): DI with 1 : 4 ratio in volume, and then 1 part of BHF (7 : 1, J.T.Baker) was added to 10 parts of the diluted HCl mixture in volume. The stock BCZT etchant can be still effective even after several months. The LSMO etchant was prepared by mixing 4 mg KI, 5 mL HCl, and 200 mL DI water, following the recipe reported elsewhere[19, 20].

Electrical *I-V* measurements were carried out on a probe station (M&M Micromanipulator) using a semiconductor analyzer (Agilent 4156C). Tungsten probes were used to directly contact exposed film after each etching step to check its electrical conductivity, and hence, to confirm the etching progress.

## Supporting Information

XRD of BCZT before and after transfer; PFM of transferred BCZT; schematics of device fabrication steps; optical microscopic images showing the optimization of BCZT wet etchant; *I-V* characteristics of SRO/BCZT/SRO device fabricated on wafer; optical microscopic images showing different device fabrication steps; schematic design of the device packaging processes for device array interconnection.

## Author declarations

**Author contributions**

H.C. and J.R. conceived the idea and designed the experiments. Y.L., H.K., and M.S. grew the samples by PLD and carried out XRD measurements. Y.L. performed PFM measurements. H.C. fabricated devices. H.C. and J.R. wrote the manuscript with input from all other authors.

**Acknowledgements**

We gratefully acknowledge support from an ARO MURI program (W911NF-21-1-0327). The growth equipment used for the materials synthesis were supported by an Air Force Office of Scientific Research grant (FA9550-22-1-0117).

**Conflict of interest**

The authors declare no competing financial interests.

# References

(1) Demeester, P.; Pollentier, I.; De Dobbelaere, P.; Brys, C.; Van Daele, P. Epitaxial lift-off and its applications. *Semicond. Sci. Technol.* **1993**, *8* (6), 1124.

(2) Cheng, C.-W.; Shiu, K.-T.; Li, N.; Han, S.-J.; Shi, L.; Sadana, D. K. Epitaxial lift-off process for gallium arsenide substrate reuse and flexible electronics. *Nat. Commun.* **2013**, *4* (1), 1577.

(3) Van Geelen, A.; Hageman, P.; Bauhuis, G.; Van Rijsingen, P.; Schmidt, P.; Giling, L. Epitaxial lift-off GaAs solar cell from a reusable GaAs substrate. *Mater. Sci. Eng., B* **1997**, *45* (1-3), 162-171.

(4) Yoon, J.; Jo, S.; Chun, I. S.; Jung, I.; Kim, H.-S.; Meitl, M.; Menard, E.; Li, X.; Coleman, J. J.; Paik, U. GaAs photovoltaics and optoelectronics using releasable multilayer epitaxial assemblies. *Nature* **2010**, *465* (7296), 329-333.

(5) Bauhuis, G. J.; Mulder, P.; Haverkamp, E. J.; Huijben, J.; Schermer, J. J. 26.1% thin-film GaAs solar cell using epitaxial lift-off. *Sol. Energy Mater. Sol. Cells* **2009**, *93* (9), 1488-1491.

(6) Gai, B.; Sun, Y.; Lim, H.; Chen, H.; Faucher, J.; Lee, M. L.; Yoon, J. Multilayer-grown ultrathin nanostructured GaAs solar cells as a cost-competitive materials platform for III–V photovoltaics. *ACS Nano* **2017**, *11* (1), 992-999.

(7) Kang, D.; Young, J. L.; Lim, H.; Klein, W. E.; Chen, H.; Xi, Y.; Gai, B.; Deutsch, T. G.; Yoon, J. Printed assemblies of GaAs photoelectrodes with decoupled optical and reactive interfaces for unassisted solar water splitting. *Nat. Energy* **2017**, *2* (5), 1-5.

(8) Dawber, M.; Rabe, K.; Scott, J. Physics of thin-film ferroelectric oxides. *Rev. Mod. Phys.* **2005**, *77* (4), 1083.

(9) Choi, K. J.; Biegalski, M.; Li, Y.; Sharan, A.; Schubert, J.; Uecker, R.; Reiche, P.; Chen, Y.; Pan, X.; Gopalan, V. Enhancement of ferroelectricity in strained BaTiO$_3$ thin films. *Science* **2004**, *306* (5698), 1005-1009.

(10) Cohen, R. E. Origin of ferroelectricity in perovskite oxides. *Nature* **1992**, *358* (6382), 136-138.

(11) Ohtomo, A.; Hwang, H. A high-mobility electron gas at the LaAlO$_3$/SrTiO$_3$ heterointerface. *Nature* **2004**, *427* (6973), 423-426.

(12) Hong, M.; Liou, S. H.; Kwo, J.; Davidson, B. Superconducting Y-Ba-Cu-O oxide films by sputtering. *Appl. Phys. Lett.* **1987**, *51* (9), 694-696.

(13) Schooley, J.; Hosler, W.; Cohen, M. L. Superconductivity in semiconducting SrTiO$_3$. *Phys. Rev. Lett.* **1964**, *12* (17), 474.

(14) Ahadi, K.; Galletti, L.; Li, Y.; Salmani-Rezaie, S.; Wu, W.; Stemmer, S. Enhancing superconductivity in SrTiO$_3$ films with strain. *Sci. Adv.* **2019**, *5* (4), eaaw0120.

(15) Yadav, A.; Nelson, C.; Hsu, S.; Hong, Z.; Clarkson, J.; Schlepütz, C.; Damodaran, A.; Shafer, P.; Arenholz, E.; Dedon, L. Observation of polar vortices in oxide superlattices. *Nature* **2016**, *530* (7589), 198-201.

(16) Lee, H.; Campbell, N.; Lee, J.; Asel, T.; Paudel, T.; Zhou, H.; Lee, J.; Noesges, B.; Seo, J.; Park, B. Direct observation of a two-dimensional hole gas at oxide interfaces. *Nat. Mater.* **2018**, *17* (3), 231-236.

(17) Novoselov, K.; Mishchenko, A.; Carvalho, A.; Castro Neto, A. 2D materials and van der Waals heterostructures. *Science* **2016**, *353* (6298), aac9439.

(18) Fan, S.; Vu, Q. A.; Tran, M. D.; Adhikari, S.; Lee, Y. H. Transfer assembly for two-dimensional van der Waals heterostructures. *2D Mater.* **2020**, *7* (2), 022005.

(19) Bakaul, S. R.; Serrao, C. R.; Lee, M.; Yeung, C. W.; Sarker, A.; Hsu, S.-L.; Yadav, A. K.; Dedon, L.; You, L.; Khan, A. I. Single crystal functional oxides on silicon. *Nat. Commun.* **2016**, *7* (1), 1-5.

(20) Bakaul, S. R.; Serrao, C. R.; Lee, O.; Lu, Z.; Yadav, A.; Carraro, C.; Maboudian, R.; Ramesh, R.; Salahuddin, S. High speed epitaxial perovskite memory on fexible substrates. *Adv. Mater.* **2017**, *29* (11).

(21) Lu, D.; Baek, D. J.; Hong, S. S.; Kourkoutis, L. F.; Hikita, Y.; Hwang, H. Y. Synthesis of freestanding single-crystal perovskite films and heterostructures by etching of sacrificial water-soluble layers. *Nat. Mater.* **2016**, *15* (12), 1255-1260.

(22) Ji, D.; Cai, S.; Paudel, T. R.; Sun, H.; Zhang, C.; Han, L.; Wei, Y.; Zang, Y.; Gu, M.; Zhang, Y. Freestanding crystalline oxide perovskites down to the monolayer limit. *Nature* **2019**, *570* (7759), 87-90.

(23) Shen, L.; Wu, L.; Sheng, Q.; Ma, C.; Zhang, Y.; Lu, L.; Ma, J.; Ma, J.; Bian, J.; Yang, Y. Epitaxial lift-off of centimeter-scaled spinel ferrite oxide thin films for flexible electronics. *Adv. Mater.* **2017**, *29* (33), 1702411.

(24) Luo, Z.-D.; Peters, J. J.; Sanchez, A. M.; Alexe, M. Flexible memristors based on single-crystalline ferroelectric tunnel junctions. *ACS Appl. Mater. Interfaces* **2019**, *11* (26), 23313-23319.

(25) Huang, J.-K.; Wan, Y.; Shi, J.; Zhang, J.; Wang, Z.; Wang, W.; Yang, N.; Liu, Y.; Lin, C.-H.; Guan, X. High-*κ* perovskite membranes as insulators for two-dimensional transistors. *Nature* **2022**, *605* (7909), 262-267.

(26) Yang, A. J.; Han, K.; Huang, K.; Ye, C.; Wen, W.; Zhu, R.; Zhu, R.; Xu, J.; Yu, T.; Gao, P. Van der Waals integration of high-$\kappa$ perovskite oxides and two-dimensional semiconductors. *Nat. Electron.* **2022**, *5* (4), 233-240.

(27) Hong, S. S.; Gu, M.; Verma, M.; Harbola, V.; Wang, B. Y.; Lu, D.; Vailionis, A.; Hikita, Y.; Pentcheva, R.; Rondinelli, J. M. Extreme tensile strain states in $La_{0.7}Ca_{0.3}MnO_3$ membranes. *Science* **2020**, *368* (6486), 71-76.

(28) Xu, R.; Huang, J.; Barnard, E. S.; Hong, S. S.; Singh, P.; Wong, E. K.; Jansen, T.; Harbola, V.; Xiao, J.; Wang, B. Y. Strain-induced room-temperature ferroelectricity in $SrTiO_3$ membranes. *Nat. Commun.* **2020**, *11* (1), 3141.

(29) Gu, K.; Katayama, T.; Yasui, S.; Chikamatsu, A.; Yasuhara, S.; Itoh, M.; Hasegawa, T. Simple method to obtain large-size single-crystalline oxide sheets. *Adv. Funct. Mater.* **2020**, *30* (28), 2001236.

(30) Zhang, B.; Yun, C.; MacManus-Driscoll, J. L. High yield transfer of clean large-area epitaxial oxide thin films. *Nano-Micro Letters* **2021**, *13*, 1-14.

(31) Singh, P.; Swartz, A.; Lu, D.; Hong, S. S.; Lee, K.; Marshall, A. F.; Nishio, K.; Hikita, Y.; Hwang, H. Y. Large-area crystalline $BaSnO_3$ membranes with high electron mobilities. *ACS Appl. Electron. Mater.* **2019**, *1* (7), 1269-1274.

(32) Chen, Z.; Wang, B. Y.; Goodge, B. H.; Lu, D.; Hong, S. S.; Li, D.; Kourkoutis, L. F.; Hikita, Y.; Hwang, H. Y. Freestanding crystalline $YBa_2Cu_3O_{7-x}$ heterostructure membranes. *Phys. Rev. Mater.* **2019**, *3* (6), 060801.

(33) Liu, W.; Ren, X. Large piezoelectric effect in Pb-free ceramics. *Phys. Rev. Lett.* **2009**, *103* (25), 257602.

(34) Xue, D.; Zhou, Y.; Bao, H.; Zhou, C.; Gao, J.; Ren, X. Elastic, piezoelectric, and dielectric properties of Ba(Zr$_{0.2}$Ti$_{0.8}$)O$_3$-50(Ba$_{0.7}$Ca$_{0.3}$)TiO$_3$ Pb-free ceramic at the morphotropic phase boundary. *J. Appl. Phys.* **2011**, *109* (5).

(35) Dupuy, A. D.; Kodera, Y.; Garay, J. E. Unprecedented electro-optic performance in lead-free transparent ceramics. *Adv. Mater.* **2016**, *28* (36), 7970-7977.

(36) Silva, J. P.; Silva, J. M.; Oliveira, M. J.; Weingärtner, T.; Sekhar, K. C.; Pereira, M.; Gomes, M. J. High-performance ferroelectric–dielectric multilayered thin films for energy storage capacitors. *Adv. Funct. Mater.* **2019**, *29* (6), 1807196.

(37) Liu, Y.; Wang, Z.; Thind, A. S.; Orvis, T.; Sarkar, D.; Kapadia, R.; Borisevich, A. Y.; Mishra, R.; Khan, A. I.; Ravichandran, J. Epitaxial growth and dielectric characterization of atomically smooth 0.5Ba(Zr$_{0.2}$Ti$_{0.8}$)O$_3$–0.5(Ba$_{0.7}$Ca$_{0.3}$)TiO$_3$ thin films. *J. Vac. Sci. Technol. A* **2019**, *37* (1), 011502.

(38) Kumarasubramanian, H.; Ravindran, P. V.; Liu, T.-R.; Song, T.; Surendran, M.; Chen, H.; Buragohain, P.; Tung, I.; Gupta, A. S.; Steinhardt, R. Kinetic control of ferroelectricity in ultrathin epitaxial barium titanate capacitors. *arXiv preprint arXiv:2407.13953* **2024**.

(39) Jiang, Y.; Parsonnet, E.; Qualls, A.; Zhao, W.; Susarla, S.; Pesquera, D.; Dasgupta, A.; Acharya, M.; Zhang, H.; Gosavi, T. Enabling ultra-low-voltage switching in BaTiO$_3$. *Nat. Mater.* **2022**, *21* (7), 779-785.

(40) Weber, D.; Vőfély, R.; Chen, Y.; Mourzina, Y.; Poppe, U. Variable resistor made by repeated steps of epitaxial deposition and lithographic structuring of oxide layers by using wet chemical etchants. *Thin Solid Films* **2013**, *533*, 43-47.

(41) Miller, R. A.; Bernstein, J. J. A novel wet etch for patterning lead zirconate-titanate (PZT) thin-films. *Integrated Ferroelectrics* **2000**, *29* (3-4), 225-231.

(42) Zhang, R.; Yang, C.; Yu, A.; Wang, B.; Tang, H.; Chen, H.; Zhang, J. Wet chemical etching method for BST thin films annealed at high temperature. *Appl. Surf. Sci.* **2008**, *254* (21), 6697-6700.

(43) Gai, B.; Geisz, J. F.; Friedman, D. J.; Chen, H.; Yoon, J. Printed assemblies of microscale triple-junction inverted metamorphic GaInP/GaAs/InGaAs solar cells. *Prog Photovolt Res Appl.* **2019**, *27* (6), 520-527.

(44) Chen, H.; Lee, S.-M.; Montenegro, A.; Kang, D.; Gai, B.; Lim, H.; Dutta, C.; He, W.; Lee, M. L.; Benderskii, A. Plasmonically enhanced spectral upconversion for improved performance of GaAs solar cells under nonconcentrated solar illumination. *ACS Photonics* **2018**, *5* (11), 4289-4295.

(45) Kang, D.; Chen, H.; Yoon, J. Stretchable, skin-conformal microscale surface-emitting lasers with dynamically tunable spectral and directional selectivity. *Appl. Phys. Lett.* **2019**, *114* (4), 041103.

(46) Yablonovitch, E.; Hwang, D.; Gmitter, T.; Florez, L.; Harbison, J. Van der Waals bonding of GaAs epitaxial liftoff films onto arbitrary substrates. *Appl. Phys. Lett.* **1990**, *56* (24), 2419-2421.

(47) Lau, J. H.; Lau, J. H. *Advanced packaging*; Springer, 2021.

(48) Gao, W.; Ota, H.; Kiriya, D.; Takei, K.; Javey, A. Flexible electronics toward wearable sensing. *Acc. Chem. Res.* **2019**, *52* (3), 523-533.

(49) Kang, J.-H.; Shin, H.; Kim, K. S.; Song, M.-K.; Lee, D.; Meng, Y.; Choi, C.; Suh, J. M.; Kim, B. J.; Kim, H. Monolithic 3D integration of 2D materials-based electronics towards ultimate edge computing solutions. *Nat. Mater.* **2023**, *22* (12), 1470-1477.

(50) Nian, L.; Sun, H.; Wang, Z.; Xu, D.; Hao, B.; Yan, S.; Li, Y.; Zhou, J.; Deng, Y.; Hao, Y. $Sr_4Al_2O_7$: A New Sacrificial Layer with High Water Dissolution Rate for the Synthesis of Freestanding Oxide Membranes. *Adv. Mater.* **2024**, 2307682.

(51) Peng, B.; Peng, R.-C.; Zhang, Y.-Q.; Dong, G.; Zhou, Z.; Zhou, Y.; Li, T.; Liu, Z.; Luo, Z.; Wang, S. Phase transition enhanced superior elasticity in freestanding single-crystalline multiferroic $BiFeO_3$ membranes. *Sci. Adv.* **2020**, *6* (34), eaba5847.

**TOC**

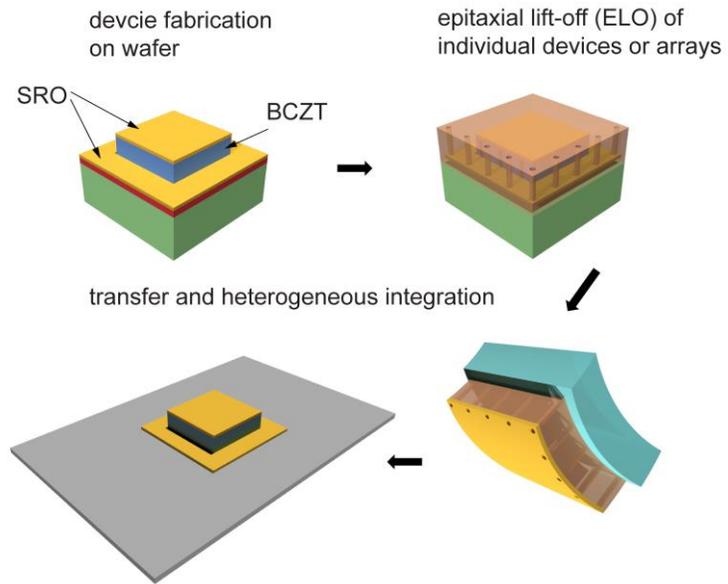

# Supporting Information

# A strategy for fabricating micro-scale freestanding single-crystalline complex oxide device arrays

Huandong Chen[1], Yang Liu[1], Harish Kumarasubramanian[1], Mythili Surendran[1,3], Jayakanth Ravichandran[1,2,3]*

[1]Mork Family Department of Chemical Engineering and Materials Science, University of Southern California, Los Angeles, California, USA

[2]Ming Hsieh Department of Electrical and Computer Engineering, University of Southern California, Los Angeles, California, USA

[3]Core Center for Excellence in Nano Imaging, University of Southern California, Los Angeles, California, USA

*Email: j.ravichandran@usc.edu

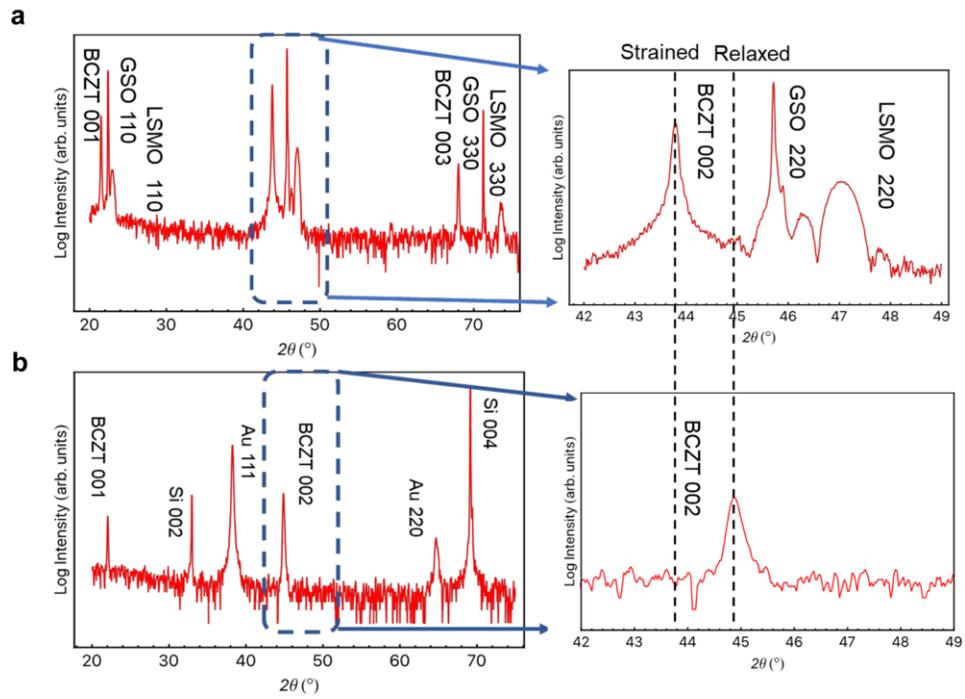

**Figure S1.** XRD scan of BCZT film (**a**) before and (**b**) after transfer. Each reflection is indexed with corresponding material. The insets on the right are the short-angle XRD scans for 42-49° near the most intense 002 reflection of BCZT.

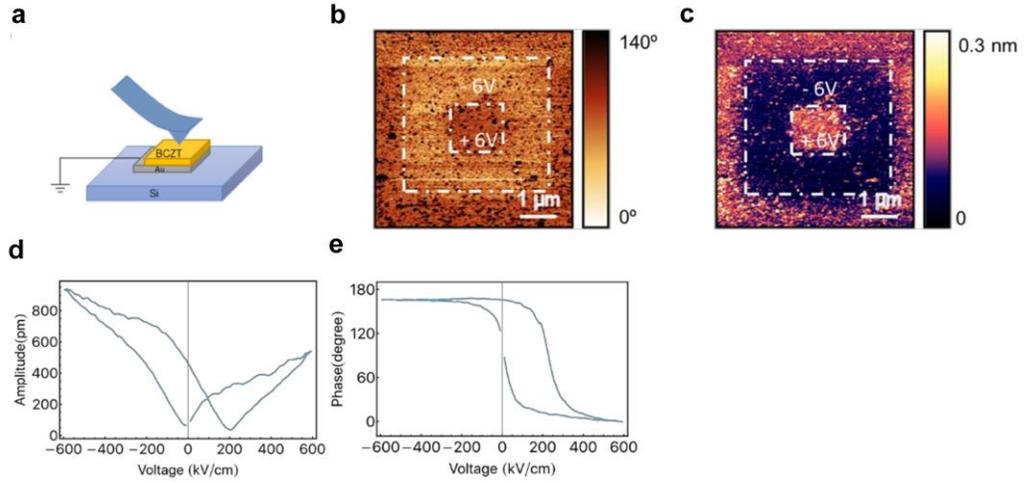

**Figure S2. PFM responses of transferred BCZT films.** (**a**) Schematic illustration of PFM setup for characterizing transferred BCZT film. The box-in-box pattern is shown in terms of phase and amplitude in (**b**) and (**c**), respectively. (**d**) and (**e**) show the amplitude and phase responses as a function of field. A 100 nm-thick BCZT film was used for the measurement and the shape of BCZT hysteresis loop illustrates the characteristic feature of a relaxor ferroelectric.

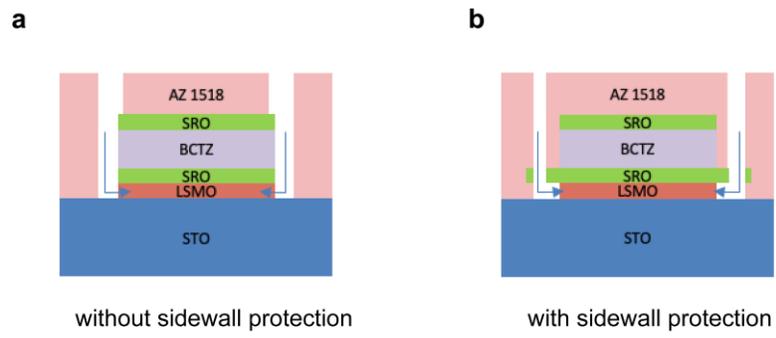

**Figure S3.** Schematics of etching holes design of (**a**) without sidewall protection, and (**b**) with sidewall protection

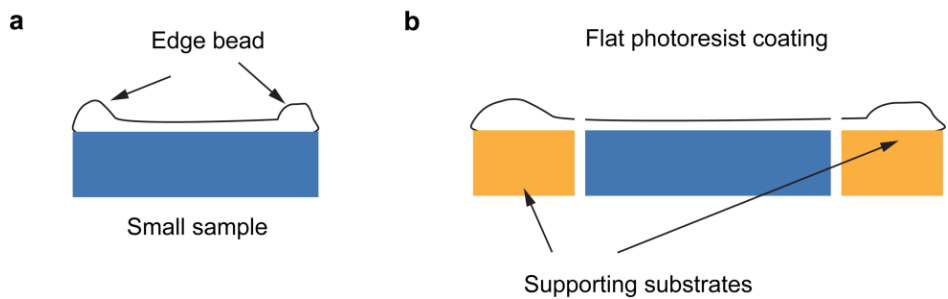

**Figure S4.** Schematics of photoresist spin-coating on small-sized PLD-grown oxide samples. (**a**) direct spin-coating on small samples without supporting substrates around, which leads to sever edge bead issues, and (**b**) spin coating on small samples with supporting substrates. Edges bead issue is minimized with supporting substrates, and hence, flat photoresist coating can be obtained. This is particularly important for fabricating devices on small pieces of oxide samples with lateral dimensions smaller than 2 mm.

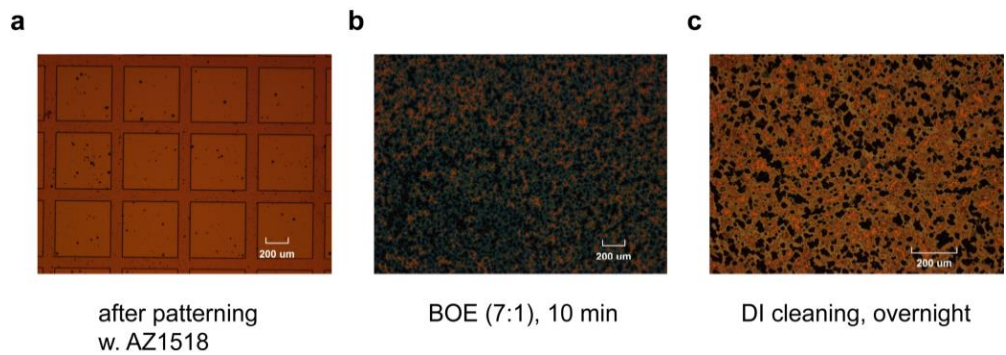

**Figure S5.** BCZT wet etching test with HF-based etchant. Optical microscopic images of (**a**) photoresist patterning on a BCZT/SRO/STO test sample, (**b**) after wert etching in BOE (7:1) for 10 min. All the patterns and oxide films are gone, leaving a large amount of etching residues and a very rough surface. (**c**) after DI cleaning overnight.

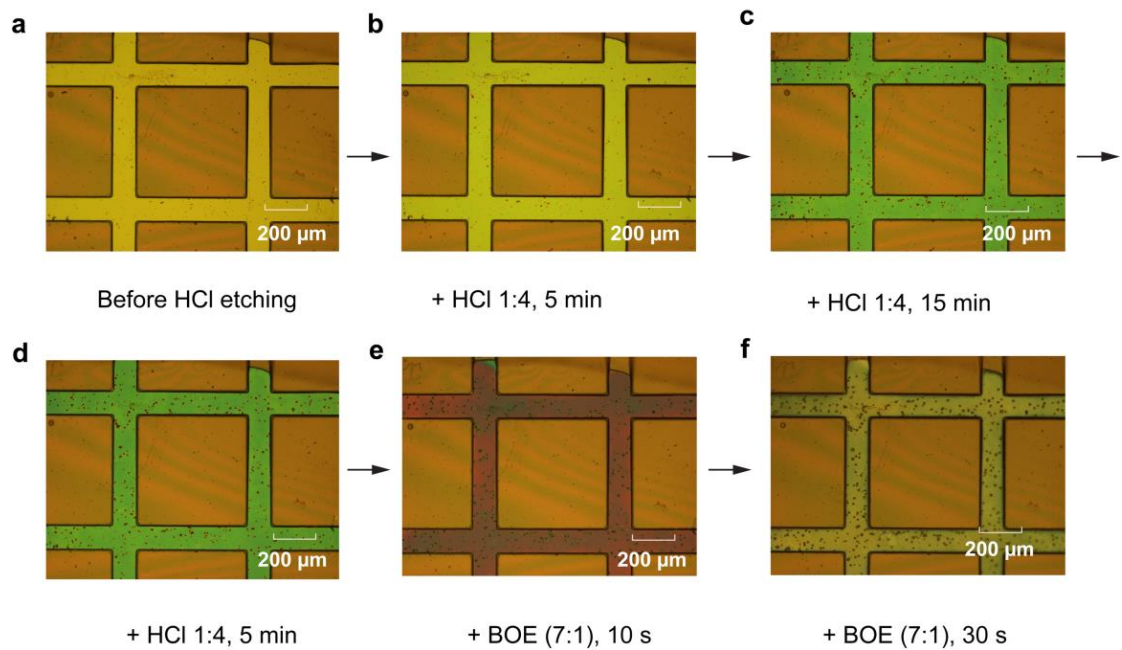

**Figure S6.** BCZT wet etching test with HCl-based etchant. (**a**) to (**d**) Optical microscopic images of a BCZT device stack before and after consecutive etching by HCl 1:4. The gradual change of color corresponds to the thinning and continuous etching of BCZT film. No discernible difference was observed between (**c**) and (**d**), which indicates that BCZT is not effectively removed in HCl-based etchant beyond certain amount of time. (**e**) and (**f**) BCZT etching continues by dipping the same piece of sample in BOE etchant for tens of seconds. Therefore, both HF and HCl contents were used to prepare the optimized BCZT etchant.

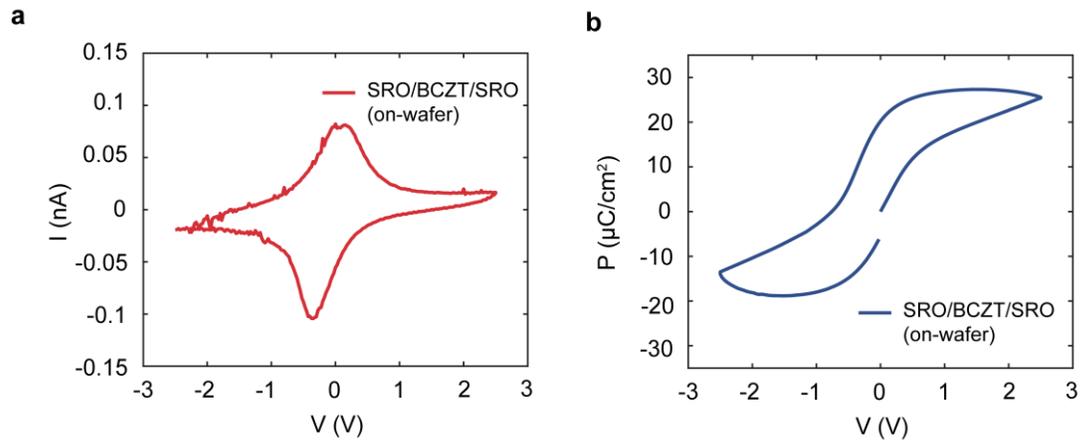

**Figure S7.** (**a**) and (**b**) Representative *I-V* and *P-V* characteristics of an SRO/BCZT/SRO capacitor fabricated on wafer.

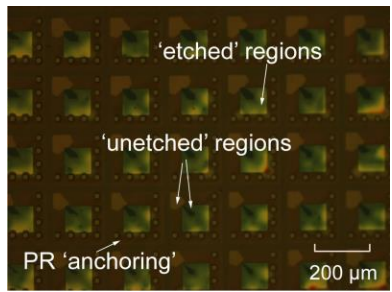

**Figure S8.** Optical microscopic image showing the undercut etching process, with 'etched' regions, 'unetched' regions, and PR 'anchoring' regions clearly labeled.

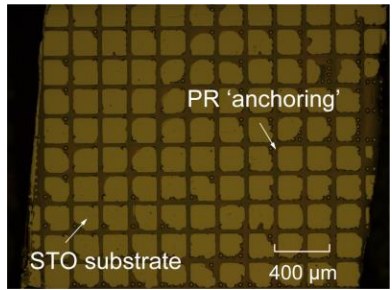

**Figure S9.** Optical microscopic image showing STO substrate after freestanding SRO/BCZT/SRO device arrays were picked up using PDMS stamp. PR 'anchoring' was left on the substrate.

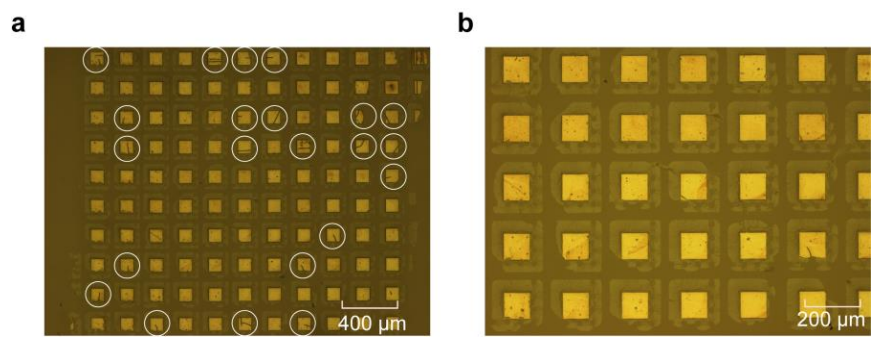

**Figure S10.** Optical microscopic images of transferred SRO/BCZT/SRO device arrays. Devices with cracks formed during the transfer process are labeled with white circles in (**a**). A zoomed-in image was shown in panel (**b**).

"VIA hole formation"      "Interconnection"

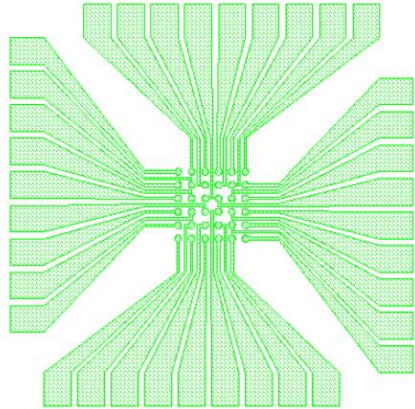

**Figure S11.** Schematic design of the device packaging processes that could potentially resolve the current electrical characterization issue on transferred capacitor arrays, including "VIA hole formation" and "metal interconnection".